# Algebraic Characterization of CNOT-Based Quantum Circuits with its Applications on Logic Synthesis

Mehdi Saeedi, Morteza Saheb Zamani, Mehdi Sedighi
*Amirkabir University of Technology, Computer Engineering Department, Tehran, Iran*
*Email: {msaeedi, szamani, msedighi}@aut.ac.ir*

**Abstract**

*The exponential speed up of quantum algorithms and the fundamental limits of current CMOS process for future design technology have directed attentions toward quantum circuits. In this paper, the matrix specification of a broad category of quantum circuits, i.e. CNOT-based circuits, are investigated. We prove that the matrix elements of CNOT-based circuits can only be zeros or ones. In addition, the columns or rows of such a matrix have exactly one element with the value of 1.*

*Furthermore, we show that these specifications can be used to synthesize CNOT-based quantum circuits. In other words, a new scheme is introduced to convert the matrix representation into its SOP equivalent using a novel quantum-based Karnaugh map extension. We then apply a search-based method to transform the obtained SOP into a CNOT-based circuit. Experimental results prove the correctness of the proposed concept.*

## 1. Introduction

It has been predicted that current CMOS technology will reach its fundamental limits in the near future[1]. On the other hand, an enormous amount of processing power is required to run many existing applications such as computer animation, molecular biology analyses, global climate and economic modeling. The demands of these applications together with the limitations of CMOS technology for increasing the processing power in the coming years have led researchers to work on new computational models.

Among various proposed computational models, quantum computing has the potential to increase the rate of advances in computing power drastically, at least for some problems [2]. In other words, there are many applications that cannot be performed on a classical Turing machine as efficiently as a quantum computer [3], [4]. The promise of exponential speed up of quantum algorithms running on quantum computers has intensified the attempts for using quantum algorithms in real world problems [3], [4].

On the other hand, a quantum algorithm needs a quantum circuit to attain its processing power. Among various quantum circuits, CNOT-based circuits have attracted more attentions in the literature (see for example [13], [5] and [12]). Due to many applications of these circuits, in this paper, the characterization of matrix representation of CNOT-based circuits is considered to be used for quantum circuit synthesis.

The rest of the paper is organized as follows: in Section 2, a brief introduction to quantum computation is presented. Previous work is reviewed in Section 3. The matrix specification of CNOT-based circuits is studied and used for quantum circuit synthesis in Section 4 and Section 5, respectively. Experimental results are reported in Section 6 and finally, Section 7 concludes the paper.

## 2. Preliminary

Quantum computation uses quantum mechanics to perform a task. A quantum bit (or qubit) is typically derived from the state of a two-level quantum system such as the ground and excited states of an atom or the vertical and horizontal polarizations of a single photon. The common notation of a qubit denotes one of these states as $|0\rangle$ and the other as $|1\rangle$. A quantum system with a collection of *n* qubits is called a quantum register of size *n*.

Unlike a classical bit which takes only one of the two pure values of 0 or 1, in quantum computation the state of a qubit can take not only two pure states $|0\rangle$ and $|1\rangle$, but also any linear combinations of these pure states, also called superposition, resulting in exponentially larger state space. In other words, the state of a qubit $\psi$ can be written as $\psi=\alpha|0\rangle+\beta|1\rangle$ where $\alpha$ and $\beta$ are complex numbers and $\alpha^2+\beta^2=1$.

Although a qubit can take any arbitrary value, when a quantum system is measured, its state collapses into the basis, i.e. $|0\rangle$ and $|1\rangle$, with the probability of $\alpha^2$ and $\beta^2$, respectively. Therefore, no information about previous superposition state remains. It is common to denote the state of a single qubit by a 2×1 vector as $[\alpha\ \beta]^T$. So, the state of a quantum register of size $n$ can also be shown by an $2^n \times 1$ vector $[\alpha_1\ \alpha_2\ \ldots\ \alpha_{2^n}]^T$ where each $\alpha_i$ ($i=1,2,..,2^n$) is a complex number and $\alpha_1^2+\alpha_2^2\ldots+\alpha_{2^n}^2=1$. If only one $\alpha_i$ ($i=1,2,..,2^n$) is set to be one, a pure quantum state of size $n$ is formed.

An $n$-qubit *quantum gate* is a device which performs a specific unitary operation on selected qubits in a specific period of time. An n-qubit quantum gate has a unitary $2^n \times 2^n$ matrix, called quantum matrix (QMatrix) in this paper, describing its functionality. A matrix $M$ is unitary if $MM^+=I$ where $M^+$ is the conjugate transpose of $M$ and $I$ is the identity matrix.

Previously, various quantum gates with different functionalities have been proposed [5]. Among them, identity (I), NOT, CNOT, C$^2$NOT and SWAP gates comprise an important class of quantum gates and often appear in the quantum computing literature [13], [5], [12]. The library which contains these gates are commonly called "CNTS" library [7], [13]. The CNTS library gates are shown in Figure 1. In this figure, control, target and contact qubits are represented as •, ⊕ and | symbols, respectively. These gates are defined as follows:

- 1-qubit identity gate (I) with matrix $M_I$ which works as a horizontal wire and may be used to construct any other quantum gates using tensor product.

- 1-qubit NOT gate with matrix $M_{NOT}$ that inverts the working qubit.

- 2-qubit CNOT gate, also called Feynman gate, with matrix $M_{CNOT}$ which works as follows: if the control qubit is $|1\rangle$, the target qubit is inverted; otherwise it is left unchanged.

- 2-qubit SWAP gate with matrix $M_{SWAP}$ that exchanges the values of its two qubits.

- 3-qubit C$^2$NOT gate, also called Toffoli gate, with matrix $M_{C2NOT}$ that works as follows: if both control qubits are $|1\rangle$, the target is inverted; otherwise it is left unchanged.

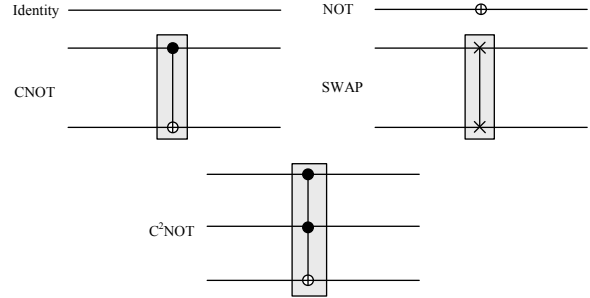

**Figure 1- Basic quantum gates used in this paper**

For a more detailed description of various quantum gates, interested readers can refer to [5]. By using these gates and tensor product, any other related gates can be constructed. For example, in Figure 2, a new compound gate and its QMatrix are shown. Two or more quantum gates can also be cascaded to construct a quantum circuit. This operation is denoted by ∘ symbol in this paper. The QMatrix of a quantum circuit is derived from its gate QMatrices using matrix multiplication. For example, in Figure 3 a quantum circuit and its QMatrix are shown.

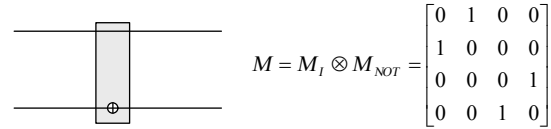

$$M = M_I \otimes M_{NOT} = \begin{bmatrix} 0 & 1 & 0 & 0 \\ 1 & 0 & 0 & 0 \\ 0 & 0 & 0 & 1 \\ 0 & 0 & 1 & 0 \end{bmatrix}$$

**Figure 2- A compound gate constructed from an identity and a NOT gate.**

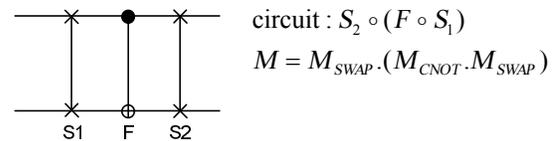

circuit : $S_2 \circ (F \circ S_1)$
$M = M_{SWAP}.(M_{CNOT}.M_{SWAP})$

**Figure 3- Cascading quantum gates to construct a quantum circuit and its QMatrix**

Positive polarity Reed-Muller (PPRM) expansion [6] in Boolean logic is used in the last part of this paper to synthesize a CNOT-based circuit. PPRM expansion is a canonical representation that uses only un-complemented (or positive) variables as follows:

$$f(x_1, x_1,..., x_n) = a_0 \oplus a_1 x_1 \oplus \cdots \oplus a_n x_n \oplus$$
$$a_{12} x_1 x_2 \oplus \cdots \oplus a_{n,n-1} x_{n-1} x_n \oplus \quad (1)$$
$$\ldots \oplus a_{12\ldots n} x_1 x_2 \cdots x_n$$

In the following section, previous work on quantum circuit synthesis is reviewed.

## 3. Previous work

Several algorithms have recently been proposed to synthesize a quantum circuit. Toffoli in [7] presented an algorithm to implement a function using CNTS library. As this algorithm uses many extra qubits, it cannot be used to synthesize a general quantum circuit efficiently. Some authors used transformation-based algorithms for quantum circuit synthesis [8]-[10]. However, these algorithms usually use local transformations to optimize the results of other algorithms. Miller in [11] considered the use of Rademacher-Walsh spectral techniques and two-place decompositions of Boolean functions to synthesize a reversible circuit. In [12], a new incremental approach was presented which uses shared binary decision diagrams for representing a reversible function and measuring circuit complexity. The proposed algorithm selects reversible gates based on the complexity of the rest of logic.

The authors of [13] investigated a number of techniques to synthesize optimal and near-optimal reversible circuits that require little or no temporary storage. They also provided some properties about even and odd permutation functions. The authors of this paper [14] proposed a search-based algorithm to find a CNOT-based implementation of a reversible functions based on PPRM expansion. In [15], an approach to synthesize a quantum circuit was proposed which uses symbolic reachability analysis where the primary inputs are assumed to be purely binary.

In [16], Shende et al. presented a top-down structure based on Cosine-Sine decomposition to introduce quantum multiplexer and used it to propose a synthesis algorithm in terms of quantum multiplexers. In [17], a recursive synthesis method for ternary quantum circuits based on the Cosine-Sine unitary matrix decomposition was presented. The authors of [18] and [19] presented an algorithm to decompose the matrix of a quantum circuit into the unitary matrices of elementary quantum gates. However, these methods are not practical to synthesize a general quantum circuit of arbitrary size. Abdollahi and Pedram [20] presented a quantum decision diagram structure and used it to synthesize quantum circuits using $R_x(\theta)$ rotation gates.

In order to reduce the CPU time of synthesis algorithms, some researchers used an evolutionary synthesis algorithm [21]. Some authors used mathematical methods to put a lower bound on the number of quantum gates required to synthesize a quantum circuit [22]. The authors of [23] proposed an approach to optimally synthesize 3-qubit quantum circuits by group theory where the primary inputs are assumed to be binary states.

As the size of a quantum circuit increases drastically, a practical algorithm for quantum circuit synthesis becomes extremely difficult. As will be shown in Section 5, we use matrix specifications to propose a systematic methodology for CNOT-based quantum circuit synthesis.

## 4. Matrix characterizations

In this section, we introduce several notations and theorems to study the matrix characterizations of CNOT-based quantum circuits.

**Notation 1:** A 3-qubit $C^2NOT$ gate is denoted as $C^2NOT_3(1,2,3)$ where two first qubits are control qubits and the last one is the target.

**Notation 2:** An n-qubit quantum gate consisting of a $C^2NOT$ gate and $n-3$ identity gates is denoted as $C^2NOT_n(i,j,k)$ where $i, j$ and $k$ are nonadjacent qubits. If $i, j$ and $k$ are adjacent, a $C^2NOT_n(i,i+1,i+2)$ gate is formed.

**Notation 3:** An n-qubit SWAP gate containing $n-2$ identity gates and a 2-qubit SWAP gate is denoted as $SWAP_n(i,i+1)$ where $i$ and $i+1$ are two adjacent qubits.

**Notation 4:** Let $U$ be any arbitrary quantum gate. We use the notation $U^{\otimes n}$ to show $n$ parallel executions of $U$ gate on $n$ adjacent qubits. The similar notation is used for its QMatrix representation.

**Definition 1:** The QMatrix of an n-qubit quantum circuit is *well-formed* if it has the following two conditions:

1. Matrix elements can only be zeros or ones.
2. Each column or row has exactly one element with a value of 1.

**Observation 1:** The QMatrix of $C^2NOT_3(1,2,3)$ is well-formed.

**Proof:** The gate $C^2NOT_3(1,2,3)$ has a $2^3 \times 2^3$ QMatrix with the following specification [5].

$$m_{ij} = \begin{cases} 1 & (1 \leq i = j \leq 6) \, or \, (i = 7, j = 8) \, or \, (i = 8, j = 7) \\ 0 & otherwise \end{cases} \quad (2)$$

Therefore, it is well-formed. □

**Observation 2:** The QMatrix of $SWAP_2(1,2)$ is well-formed.

**Proof:** The gate $SWAP_2(1,2)$ has a $2^2 \times 2^2$ QMatrix with the following specification [5].

$$M_{SWAP} = \begin{bmatrix} 1 & 0 & 0 & 0 \\ 0 & 0 & 1 & 0 \\ 0 & 1 & 0 & 0 \\ 0 & 0 & 0 & 1 \end{bmatrix} \quad (3)$$

Therefore, it is well-formed. □

**Theorem 1:** The set of well-formed matrices is closed under tensor product ($\otimes$) operation.

**Proof:** Let two well-formed matrices $A$ and $B$ be of dimensions $D_A \times D_A$ and $D_B \times D_B$, respectively. Furthermore, suppose that a matrix $C$ is defined as $C = A \otimes B$. Therefore, $C$ is a $(D_A * D_B) \times (D_A * D_B)$ matrix. The elements of $A$, $B$ and $C$ are represented as $a_{i,j}$, $b_{i,j}$ and $c_{i,j}$, respectively. Based on the definition of tensor product, we have:

$$c_{i,j} = a_{p,q} * b_{i-(p-1)D_B, j-(q-1)D_B} \\ (p-1)D_B < i \leq pD_B, \\ (q-1)D_B < j \leq qD_B, 0 \leq p, q \leq D_A \quad (4)$$

Since $A$ is a well-formed matrix, each of its elements can only be 0 or 1. If $a_{i,j}$ is zero, $c_{i,j}$ becomes zero, otherwise it becomes $b_{i-(p-1)D_B, j-(q-1)D_B}$ which also gets 0 or 1 (recall that $B$ is well-formed). Therefore, condition 1 of Definition 1 holds. On the other hand, Equation (5) shows the elements of the $I^{th}$ row of matrix $C$.

$$c_{I,j} = a_{p,q} * b_{I-(p-1)D_B, j-(q-1)D_B} \\ (q-1)D_B < j \leq qD_B, 0 \leq p, q \leq D_A \quad (5)$$

Since each row of matrix $A$ has exactly one element with a value of 1, it can be seen that $a_{1,q}$ ($0 \leq q \leq D_A$) is one for exactly one $q$. Therefore, Equation (5) is reduced to Equation (6).

$$if \, (k-1) * D_B < j \leq k * D_B \, then \\ c_{I,j} = a_{1,k} * b_{I, j-(k-1)D_B} \, else \, c_{I,j} = 0 \quad (6)$$

Alternatively, $b_{I, j-(k-1)D_B}$ shows the elements of the $I^{th}$ row of matrix $B$ for $(k-1) * D_B < j \leq k * D_B$. As $B$ is well-formed, only one of these elements is 1. Thus $b_{I, j-(k-1)D_B}$ is 1 for an arbitrary value of $j$, such as $J$. Therefore, among the elements of the $I^{th}$ row of matrix $C$, *only $c_{I,J}$* is equal to 1. □

**Corollary 1:** The QMatrix of $C^2NOT_n(i,i+1,i+2)$ is well-formed.

**Proof:** Based on Notation 2 and Notation 4, we have:

$$C^2NOT_n(i, i+1, i+2) = \\ I^{\otimes(i-1)} \otimes C^2NOT_3(i, i+1, i+2) \otimes I^{\otimes(n-i-2)} \quad (7)$$

Therefore, its QMatrix can be written as Equation (8)

$$M_{C^2NOT_n(i,i+1,i+2)} = \\ M_{I^{\otimes(i-1)}} \otimes \left( M_{C^2NOT_3(i,i+1,i+2)} \otimes M_{I^{\otimes(n-i-2)}} \right) \quad (8)$$

It can be seen that the QMatrix of an n-qubit identity gate is a $2^n \times 2^n$ identity matrix. Therefore, $I^{\otimes(i-1)}$ and $I^{\otimes(n-i-2)}$ gates have identity matrices of sizes $2^{i-1} \times 2^{i-1}$ and $2^{n-i-2} \times 2^{n-i-2}$, respectively. Based on Theorem 1 and Observation 1, it can be seen that the matrix $\left( M_{C^2NOT_3(i,i+1,i+2)} \otimes M_{I^{\otimes(n-i-2)}} \right)$ is well-formed. Similar results can be obtained for $M_{C^2NOT_n(i,i+1,i+2)}$. □

**Corollary 2:** The QMatrix of $SWAP_n(i,i+1)$ is well-formed.

**Proof:** Based on Notation 3 and Notation 4, we have:

$$SWAP_n(i, i+1) = \\ I^{\otimes(i-1)} \otimes SWAP_2(i, i+1) \otimes I^{\otimes(n-i-1)} \quad (9)$$

Therefore, the QMatrix of this gate can be written as

$$M_{SWAP_n(i,i+1)} = \\ M_{I^{\otimes(i-1)}} \otimes \left( M_{SWAP_2(i,i+1)} \otimes M_{I^{\otimes(n-i-1)}} \right) \quad (10)$$

Since each matrix of Equation (10) is well-formed, Theorem 1 directly results in the corollary. □

**Theorem 2:** The set of well-formed matrices is closed under matrix multiplication.

**Proof:** Assume that two n×n matrices $A$ and $B$ are well-formed. Furthermore, let matrix $C$ be $C = A*B$. The elements of $A$, $B$ and $C$ are represented as $a_{i,j}$, $b_{i,j}$, and $c_{i,j}$, respectively. As $c_{i,j} = \sum a_{i,k} b_{k,j}$, specifically for the $J^{th}$ column, $c_{i,J} = a_{i,1} b_{1,J} + a_{i,2} b_{2,J} + \ldots + a_{i,n} b_{n,J}$. Since $B$ is well-formed, only one of the elements of the $J^{th}$ column is one. Therefore, it can be said that $c_{iJ}=a_{iJ}$ which gets only two possible values 0 and 1. In Other words, condition 1 of Definition 1 holds.

In order to show the correctness of condition 2, consider the $J^{th}$ column of matrix $C$

$$\begin{aligned} c_{1,J} &= a_{1,1}b_{1,J} + a_{1,2}b_{2,J} + a_{1,3}b_{3,J} + \ldots + a_{1,n}b_{n,J} \\ c_{2,J} &= a_{2,1}b_{1,J} + a_{2,2}b_{2,J} + a_{2,3}b_{3,J} + \ldots + a_{2,n}b_{n,J} \\ &\vdots \\ c_{n,J} &= a_{n,1}b_{1,J} + a_{n,2}b_{2,J} + a_{n,3}b_{3,J} + \ldots + a_{n,n}b_{n,J} \end{aligned} \quad (11)$$

As $B$ is well-formed (i.e. $b_{IJ}$ is 1 and the other elements are zero), Equation (11) is reduced to equation (12)

$$c_{1,J} = a_{1,I}, c_{2,J} = a_{2,I}, \cdots, c_{n,J} = a_{n,I} \quad (12)$$

Therefore, exactly one of the elements of $J^{th}$ column is 1. □

**Lemma 1:** Gate $C^2NOT_n(i,j,k)$ can be written as a sequence of three gates $SWAP_n(i,i+1)$, $C^2NOT_n(i+1,j,k)$ and $SWAP_n(i+1,i)$ if $j\neq i+1$.

$$\begin{aligned} C^2NOT_n(i,j,k) &= SWAP_n(i,i+1) \\ &\circ C^2NOT_n(i+1,j,k) \circ SWAP_n(i+1,i) \end{aligned} \quad (13)$$

where symbol $\circ$ shows the sequence of time.

**Proof:** As gate $SWAP_n(i,i+1)$ exchanges the values of qubits $i$ and $i+1$, and gate $SWAP_n(i+1,i)$ retrieves them back to their original values, qubits $i$ and $i+1$ are left unchanged. In addition, the $C^2NOT$ gate is used to get qubits $i+1$ and $j$ as the controls to work on qubit $k$ as the target while qubit $i+1$ takes the value of qubit i. Figure 4 shows these two equivalent circuits. □

**Theorem 3:** The QMatrix of $C^2NOT_n(i,j,k)$ is well-formed.

**Proof:** Based on the previous lemma, the following equations can be approved:

$C^2NOT(t,j,k)=SWAP_n(t,t+1)\circ C^2NOT(t+1,j,k)\circ SWAP_n(t+1,t)$ for $\forall\ t\in[i,j-2]$ and

$C^2NOT(j-1,j,t)=SWAP_n(t-1,t)\circ C^2NOT(j-1,j,t-1)\circ SWAP_n(t,t-1)$ for $\forall\ t\in[k,j+2]$.

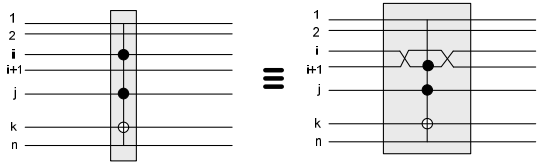

**Figure 4- Gate $C^2NOT_n(i,j,k)$ and its equivalent circuit**

Therefore, we have:

$$\begin{aligned} &C^2NOT_n(i,j,k) = \\ &SWAP_n(i,i+1)\circ\ldots\circ SWAP_n(j-2,j-1)\circ \\ &SWAP_n(k-1,k)\circ\ldots\circ SWAP_n(j+1,j+2)\circ \\ &C^2NOT(j-1,j,j+1)\circ \\ &SWAP_n(j+2,j+1)\circ\ldots\circ SWAP_n(k,k-1)\circ \\ &SWAP_n(j-1,j-2)\circ\ldots\circ SWAP_n(i+1,i) \end{aligned} \quad (14)$$

Direct using of Equation (14), Corollary 1, Corollary 2 and Theorem 2 proves Theorem 3. □

**Corollary 3:** The QMatrix of a circuit containing only $C^2NOT_n(i,j,k)$ gates is well-formed.

**Proof:** Direct using of Theorem 1, Theorem 2 and Theorem 3 proves Corollary 3. □

**Theorem 4:** If an input pure quantum state is used for a circuit with a well-formed matrix, then the produced output state is also a pure state.

**Proof:** Let the input state be $U=[u_1\ u_2\ \ldots\ u_n]^T$ and the produced output state be denoted as $\Psi=[\psi_1\ \psi_2\ \ldots\ \psi_n]^T$. If the QMatrix of the circuit is denoted as $A=[a_{i,j}]$; each element of $\Psi$ can be written as $\psi_i=\sum_k a_{i,k}u_k$. Since $U$ is a pure state, only one $u_i$ is 1 (i.e. $u_K$) and therefore, $\psi_i=a_{i,K}$. Similarly, as each column of $A$ has exactly one 1, $a_{i,k}$ will be 1 for only one $i$ $(0\leq i\leq n)$ and therefore, $\psi_I$ is 1 for only one $i$. Thus, the produced output state is a pure state. □

**Fact 1:** $C^2NOT_n(i,j,k)$ gate is a universal reversible logic gate, which means that any reversible circuit can be constructed by $C^2NOT_n(i,j,k)$ gates [7].

**Corollary 4:** The QMatrix of a reversible circuit is well-formed.

**Proof:** Direct using of Fact 1 and previous theorems leads to Corollary 4. □

We proved that a broad type of circuits (including reversible circuits) can be described by a matrix containing only zero and one elements. Furthermore, we showed that this matrix contains exactly a one in each column or row. In the rest of the paper, we use these results to transform a QMatrix into a CNOT-based quantum circuit.

## 5. Quantum circuit synthesis

As designing a large quantum circuit cannot be done manually and it requires a systematic methodology to cover various design stages, working on automatic synthesis methods for quantum circuit synthesis has received significant attention recently [8]-[23].

However, unlike Boolean logic circuits, quantum information processing is in preliminary state and no mature synthesis method for quantum circuit synthesis has been proposed yet. In order to address the problem, in the rest of this paper, we use the proposed characterizations to introduce a systematic methodology for quantum circuit synthesis.

Roughly speaking, we are interested in a quantum circuit design methodology similar to that used in conventional Boolean logic circuit designs in the sense that a higher level description is progressively converted to the final implementation. In the conventional methodology, a textual description may be converted into its equivalent truth table. Then, a canonical SOP (or POS) representation is built and optimized based on various parameters using a technology-independent optimization algorithm. In the following paragraphs, we first introduce a *quantum Karnaugh map (QKmap)* concept to transform a QMatrix into its SOP representation, and then use a search-based method to convert an SOP representation into a CNOT-based circuit.

**Definition 2:** A *quantum Karnaugh map (QKmap)* is a table to represent an n-qubit quantum circuit. The rows and columns of a QKmap are labeled using n-qubit pure quantum register values. The rules of classical Karnaugh map can be used to simplify a quantum circuit using QKmap. Figure 5 shows the QKmap for a CNOT gate.

As shown in Figure 5, the QMatrix of a CNOT gate is used to produce a pure output quantum state form a pure input quantum state. Using the QKmap, we can extract the SOPs of a CNOT gate *(a, a⊕b)*.

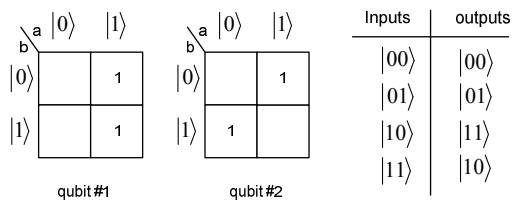

**Figure 5- QKmap for CNOT gate**

Our algorithm for QMatrix to SOP transformation is shown in Figure 6. As shown in this figure, a well-formed matrix can be transformed into its equivalent SOP. We then use this SOP to find its CNOT-based implementation.

Until now, we transformed a matrix description into an SOP representation using the method presented in the previous sections. Consequently, by using an optimization algorithm such as [24], an EXOR-Sum-of-Products (ESOP) representation can be obtained.

Then, the ESOP description will be transformed into the PPRM format by replacing *a'* with *a⊕1* for each complemented variable *a* and finally, a search-based method (such as [14]) is used to find a CNOT-based circuit for the constructed PPRM expansions (see Figure 7 for more details).

---

Matrix to SOP Transformation Algorithm

Input of the algorithm: A well-formed matrix representing a quantum circuit.
Outputs of the algorithm: An SOP representations of the input matrix in terms of working qubits.

1- As the input matrix is well-formed, it produces pure quantum states for pure input states. So, construct an intermediate table containing input and output pure states for the input matrix.
2- Using the produced intermediate table, fill the quantum Karnaugh map (QKmap).
3- By using the rules of classical Karnaugh map, simplify the QKmap.
4- Extract an SOP form for each qubit using the simplified QKmap.

**Figure 6- QMatrix to SOP transformation**

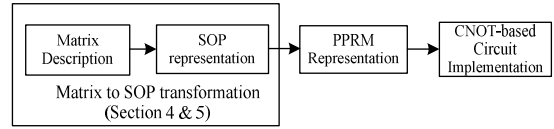

**Figure 7- QMatrix to CNOT-based circuit transformation**

It is worthwhile to note that the proposed characterizations together with the suggested flow lead to a systematic methodology for CNOT-based quantum circuit synthesis which transforms a higher level description (i.e. matrix level) into a lower level representation (i.e. gate level). In the following section, we show several examples to prove the proposed concept.

## 6. Experimental results

In this section, our experimental results for quantum circuit synthesis are presented. We use the following notation for an $n \times n$ QMatrix representation:

**Notation 5:** The QMatrix *A* is denoted as $A(x_1, x_2, \ldots, x_n)$ where $x_i$ $(i \in [1,n])$ is the column number of an element with a value of 1 in the $i^{th}$ row. For example, the QMatrix $M_{CNOT}$ is represented as *CNOT(1,2,4,3)*.

**Example 1:** Consider a 2-qubit comparator with two inputs / two outputs where the first output is the result of the comparison and the second one is a garbage output. Since, a 2-input comparator is an irreversible function, one garbage output is added to make it reversible. A possible truth table for this reversible comparator is shown in Figure 8(a). Based on this figure, the QMatrix of this circuit is (3,2,1,4). Therefore, its SOP and ESOP representations are *(a'b'+ab,b)*, *(1⊕a⊕b,b)*, respectively. Figure 8(b) shows the circuit.

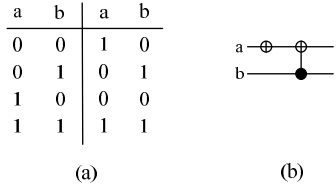

**Figure 8- A possible truth table for a 2-qubit comparator (a) and its implementation (b)**

**Example 2:** Consider a 3-qubit *Control-Inverter-Wire (CIW)* gate which flips the second qubit based on the state of the first one. The truth table of this gate is shown in Figure 9(a). The QMatrix of this gate and its SOP (ESOP) representations are *CIW(1,2,3,4,7,8,5,6)* and *(a,a⊕b,c)*, respectively. It is easy to check that Figure 9(b) describes the circuit.

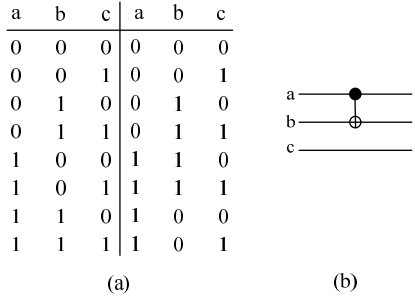

**Figure 9- The truth table of a CIW gate (a) and its CNOT-based implementation (b)**

**Example 3:** Consider a 3-qubit full-adder with three inputs ($c_{in}$, b, a) and three outputs ($g_1$, $g_2$, s) where the first two outputs (i.e. $g_1$ and $g_2$) are garbage and the last one is the result. Figure 10(a) shows a possible truth table for this function. Using this truth table will lead to the following QMatrix: *(1,8,6,7,2,3,5,4)*. Therefore, its SOP and ESOP representations are *($C_{in}$'b+ab', b, $C_{in}$⊕a⊕b)* and *(a⊕b⊕ab⊕b$C_{in}$,b,$C_{in}$⊕a⊕b)*, respectively. The circuit diagram of this example is shown in Figure 10(b).

**Example 4:** A 3-qubit majority gate outputs a 1 as its first output when two or more inputs are 1; otherwise it outputs 0. In order to make it reversible, two garbage outputs are added to this gate. Figure 11(a) shows the truth table of this gate. The QMatrix of this gate and its SOP representations are (1,2,3,5,4,6,7,8), (ab+bc+ac, ab+bc'+ ac', ac'+b'c), respectively. Figure 11(b) shows the circuit of this example.

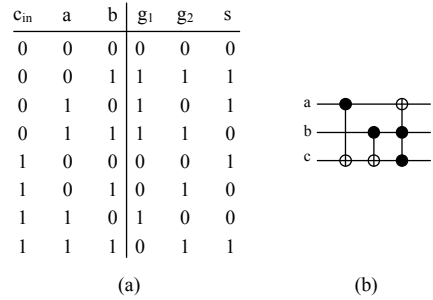

**Figure 10- A possible truth table for a 3-qubit full-adder (a) and its implementation (b)**

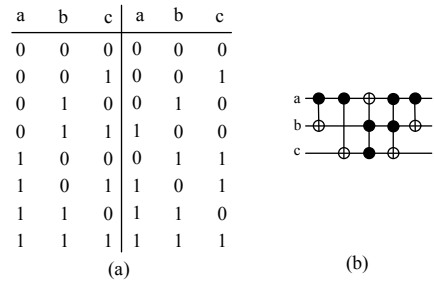

**Figure 11- The truth table of a 3-qubit majority gate (a) and its CNOT-based implementation (b)**

**Example 5:** Consider a 4-qubit swap gate with QMatrix (1,3,2,4,9,11,10,12,5,7,6,8,13,15,14,16). The SOP representations of this gate are (b,a,d,c). Figure 12 shows the circuit of this example.

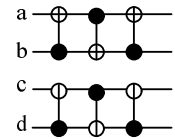

**Figure 12- The CNOT-based implementation of a 4-qubit swap gate**

**Example 6:** Consider a random 4-qubit reversible gate with QMatrix (12,4,10,3,8,14,16,15,9,2,5,11,1,13, 7,6). The proposed algorithm results in the following SOP representations:

(a'b'd'+c'd'b'+c'db+a'bc+ab'cd,a'b+bd+cd'a,a'b'c'+ c'd'a'+a'cd+b'cd+bcd', a'c'+a'd'+c'db'+abcd)

Figure 13 shows the circuit of this example.

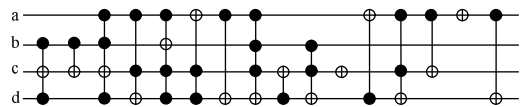

**Figure 13- The CNOT-based implementations of Example 6**

## 7. Conclusions and future works

Quantum information processing is in the pioneering stage and there is no fully optimized method for quantum circuit synthesis. In this paper, we studied the specification of CNOT-based matrices to propose a new systematic methodology for quantum circuit synthesis. In other words, by using the proposed specifications and the quantum Karnaugh map concept, we introduced an algorithm for matrix to SOP transformation. Furthermore, we used a search-based method to transform the constructed SOP into a quantum circuit.